\documentclass{chjaa}                  

\usepackage{graphicx,times}             
\usepackage{rotating}
\input{epsf.sty}                        
\input{psfig.sty}

\newcommand{\al}{\langle}
\newcommand{\ar}{\rangle}

\begin{document}

   \title{Efficient Turbulent Compressible Convection in the Deep Stellar Atmosphere}

   \volnopage{Vol.0 (200x) No.0, 000--000}      
   \setcounter{page}{1}           

   \author{Chun-Lin Tian
      \inst{1,2}\mailto{}
   , Li-Cai Deng\inst{1}, Kwing-Lam Chan\inst{3} \and Da-Run
   Xiong\inst{4}
      }

   \institute{National Astronomical Observatories, Chinese Academy of Sciences,
             Beijing 100012, China\\
             \email{cltian@bao.ac.cn}
        \and
             Graduate University of Chinese Academy of Sciences,
             Beijing 100049, China\\
        \and
             Department of Mathematics,
        Hong Kong University of Science and Technology, Hong Kong, China\\
        \and
             Purple Mountain Observatory, Chinese Academy of Sciences,
             Nanjing 210008, China\\
          }

   \date{Received~~2008 month day; accepted~~2008~~month day}

   \abstract{This paper reports an application of gas-kinetic BGK scheme
   to the computation of turbulent compressible convection in the stellar
interior. After incorporating the Sub-grid Scale (SGS) turbulence
model into the BGK scheme, we tested the effects of numerical
parameters on the quantitative relationships among the thermodynamic
variables, their fluctuations and correlations in a very deep,
initially gravity-stratified stellar atmosphere. Comparison
indicates that the thermal properties and dynamic properties are
dominated by different aspects of numerical models separately. An
adjustable Deardorff constant in the SGS model $c_\mu=0.25$ and an
amplitude of artificial viscosity in the gas-kinetic BGK scheme
$C_2=0$
  are appropriate for current study.
  We also calculated the
density-weighted auto- and cross-correlation functions in Xiong's
(\cite{xiong77}) turbulent stellar convection theories based on
which the gradient type of models of the non-local transport and the
anisotropy of the turbulence are preliminarily studied. No universal
relations or constant parameters were found for these models.
   }

   \authorrunning{Chun-Lin Tian et al. }            
    \titlerunning{Turbulent Convection in the Stellar Atmosphere}
   \maketitle

%
%
\section{Introduction}           

\label{sect:intro} Turbulent convection has tight relations to the
unsolved problems in the theory of stellar structure and evolution,
especially, for the massive stars (Deng et al. \cite{deng96a};
\cite{deng96b} and references therein). These problems cannot be
settled with pure analytical ways. Along with the development of
computing science, numerical simulations become a powerful tool to
investigate the hydrodynamic properties of
 astrophysical flows.  It is widely used in the study of formation of cluster,
accretion disk  and evolution of galaxies.  The convection in
stellar interior have also been studied by many authors with
numerical experiments. Due to the difficulties in this problem, the
progress made in this field is limited. However, it is generally
believed that numerical testing of some analytical models and local
high-resolution simulations can make a lot of sense to our
understanding of stellar convection.  So far, the numerical
hydrodynamic scheme applied to the stellar convection are
Lax-Wendroff scheme (Graham~\cite{graham75}, etc.), alternating
direction implicit method on staggered mesh (ADISM) (Chan et
al.~\cite{chan82}; \cite{chan86}, etc.), pseudo-spectrum scheme
(Hossain \& Mullan~\cite{hossain90}; \cite{hossain91}, etc.),
piecewise-parabolic method (PPM) (Porter \&
Woodward~\cite{porter94}; \cite{porter2000}, etc.), upwind scheme
and so on. At the present time, the most suitable numerical scheme
for the turbulent flow is the spectrum method, but it cannot handle
the discontinuity in the motion of fluids.

Gas-kinetic BGK scheme is a recently matured method for
computational fluid dynamics and attached much attention in the
practical problems (Xu~\cite{xu2001}). It is accurate and robust for
computing \emph{supersonic unsteady} flows. However its application
to astrophysical flows is not popular attributed to its complex.
Theoretical analysis shows that near the surface of stellar
envelope, the motion of fluids becomes supersonic which cannot be
self-consistently treated by traditional mixing-length theory (MLT)
(Deng \& Xiong~\cite{deng2001}).  Our original aim is to simulate
the supersonic turbulent convection in the outer region of yellow
giants by gas-kinetic BGK scheme which would involve many efforts in
different directions. We have already extend the BGK scheme to
include the gravitational acceleration (Tian et
al.~\cite{tian2007}). Before using it to compute the supersonic
stellar convection, the turbulence model,
 radiation transfer model and realistic input physics must be correctly implemented.

Restricted by the capacity of digital computer, we cannot afford the
very high resolution numerical experiments for the stellar type of
turbulent convection. Large eddy simulation (LES) which calculates
the large eddies explicitly while mimics the sub-grid eddies by
models may be the most
 feasible way in current stage. In current paper we implement the
SGS turbulence model (Smagorinsky~\cite{smagorinsky63};
Deardorff~\cite{deardorff71}) into the BGK scheme and validate the
three-dimensional BGK code by
 calculating the turbulent compressible convection in a deep stellar atmosphere.
For very high Reynolds number, the behaviors of the turbulent flows
are
 greatly affected by the numerical and physical
dissipation in the scheme. A investigation of these effects is very
necessary before the code is applied to the practice. By varying the
Deardorff number in the SGS model and artificial viscosity parameter
 introduced to capture the shock,  we constructed three models which
are similar to those studied by Chan \& Sofia~(\cite{chan89},
hereafter CS89; \cite{chan96}, hereafter CS96).  The empirical
relations derived by them were re-examined.
 A study of density-weighted auto- and cross-correlation function, anisotropy of
turbulence and
diffusive type of models of non-local turbulent transports in the
turbulent stellar convection theory of Xiong~(\cite{xiong77}) was conducted too.

In the next section, we give a description of gas-kinetic BGK scheme
and mainly focus on the incorporation of SGS model. The computed
physical models are formulated in Sect.~\ref{sect:model}. The
numerical results are shown in Sect.~\ref{sect:analysis} where the
discussions are also presented. The conclusions are summarized in
the last section.


\section{Gas-kinetic BGK Scheme}

\label{sect:bgk}
General numerical method for hydrodynamic problems
is to directly discretize the Navier-Stokes equations,
\begin{eqnarray}
\label{ns1}
\partial\rho/\partial t &=& -\nabla \cdot \rho \vec{v},\\
\label{ns2}
\partial\rho \vec{v}/\partial t &=& -\nabla \cdot \rho \vec{v}\vec{v}
-\nabla p+\nabla\cdot\vec{\Sigma}+\rho\vec{g},\\
\label{ns3}
\partial E/\partial t &=&-\nabla\cdot[(E+p)\vec{v}-\vec{v}\cdot\vec{\Sigma}
+\vec{F}_{d}]+\rho \vec{v}\cdot\vec{g},
\end{eqnarray}
where $\rho$ is the density, $\vec{v}$ is the velocity, $p$ is the
pressure, $\vec{g}$ is the gravitational acceleration and $E$ is the
summation of internal energy and kinetic energy.
\begin{displaymath}
\vec{\Sigma}=2\mu\vec{\sigma}+\varsigma(\nabla\cdot\vec{v})\vec{I}
\end{displaymath}
is the viscous stress tensor, where $\vec{\sigma}$ is the strain
rate tensor, $\vec{I}$ is the identity tensor, $\mu$ and $\varsigma$
are the dynamical and bulk viscosity coefficient, respectively.
$\vec{F}_d$ is the diffusive type of energy flux. Differently, the
BGK scheme works on the BGK equation,
\begin{equation}
 \label{bgk}
 \frac{\partial f}{\partial t}+\vec{c}\cdot\nabla
{f}+\vec{g}\cdot\nabla_{\vec{c}}f =\frac{f^{eq}-f}{\tau},
\end{equation}
which is an approximation of Boltzmann equation (Bhatnagar et
al.\cite{bhatnagar1954}).
 In above expression,
 $f(\vec{x},\vec{c},t)$ is the gas distribution function in the phase space,
$\vec{c}$ is the particle velocity, $\tau$ is the collision time,
and $\nabla_{\vec{c}}=(\partial/\partial c_1,\partial/\partial
c_2,\partial/\partial c_3)$. The right-hand side of equation
(\ref{bgk}) is the so-called relaxation model, which is a
simplification of the complicated collision term in the Botlzmann
equation (Vincent et al.~\cite{vincent}). It physically means that
the initially non-equilibrium distribution $f$ will approach the
equilibrium state $f^{eq}$ after the particles collide once. A
larger $\tau$ corresponds to a further state $f$ from $f^{eq}$,
i.e., the stronger non-equilibrium transport effects, such as
viscosity and conduction. In our study, the equilibrium state,
$f^{eq}$, in equation (\ref{bgk}) is taken to be Maxwellian
distribution. It can be proved mathematically  that the solutions of
Navier-Stokes equations (\ref{ns1}--\ref{ns3}) are automatically
obtained through solving the BGK equation with the following
definitions of dissipative coefficients:
\begin{equation}
\label{ntc} \mu=\tau p,\qquad
\varsigma=\frac{2}{3}\frac{\mathcal{N}}{\mathcal{N}+3}\tau p,\qquad
\kappa=\frac{\mathcal{N}+5}{2}\frac{k}{m}\tau p,
\end{equation}
where $\kappa$ is the thermal conductivity, $\mathcal{N}$ is the
internal degree of freedom for particles, $m$ is the molecule mass
and $k$ is the Boltzmann constant.

The basic idea of BGK method is to find a local approximation to the
non-linear equation~{(\ref{bgk})}. Then evaluate the macro
quantities (e.g., fluxes) using the micro distribution function $f$.
Finally, the cell average values are updated according to the
conservation laws (finite volume method). In our first attempt, the
BGK method (Xu~\cite{xu2001}) has been extended to include the
external force.  A detailed description of the three dimensional
multidimensional gas-kinetic BGK scheme for the Navier-Stokes
equations under gravitational fields was given by Tian et
al.~(\cite{tian2007}).  Here, we only outline the new implements,
i.e., the incorporation of turbulence model.

In the study of convection, the combined effects of viscosity, heat conduction and
temperature difference on the instability of flows are measured by
Rayleigh number: $Ra=g\alpha\Delta T d^3/(\nu\kappa)$, where $\alpha$ is the
thermal expansion coefficient and $\nu$ is the kinematic viscosity.
For the polytropic gas defined in Sect.~\ref{sect:model}, the
Rayleigh number can be written as:
\begin{eqnarray}
\label{rn}
 Ra=\frac{PrRT_t\mathcal{Z}^2d^2\rho_t^2}{\mu^2}\left[\frac{1-(\gamma-1)n}{\gamma}\right](n+1),
 \end{eqnarray}
where $R$ is the gas constant, $d$ is the depth of computational
domain and $\gamma$ is the ratio of specific heat, the meaning of
other symbols can be found in Sect.~\ref{sect:model}. In the
gas-kinetic BGK scheme, the viscosity is controlled by collisions of
particles. We can relate the Rayleigh number to the collision time
by the following way. Suppose during each time-step $\Delta t$, the
particles collide $\beta$ times. Then we have
\begin{equation}
\label{nts}
\Delta t=\beta \tau=\frac{\delta  \Delta x}{(c_s+v)},
\end{equation}
where $\delta$ is Courant number, $\Delta x$ is the spatial resolution,
$v$ is speed of fluids and $c_s$ is the sound speed.
From equation~(\ref{ntc}), (\ref{rn}) and (\ref{nts}),
we get
\begin{eqnarray}
 Ra \approx\frac{N^2Pr\mathcal{Z}^2(1+Ma)^2}{\delta^2}\left[1-(\gamma-1)n\right](n+1)\beta^2,
 \end{eqnarray}
where $Ma=v/c_s$ is the Mach number, $N$ is the vertical grids size
and $Pr$ is Prandtl number. For current study, we have $N\sim 50$,
$Ma\sim 1$, $\mathcal{Z}=15$, $\delta=0.3$,  $\gamma=5/3$ and
$n=0.999/(\gamma-1)$. In efficient turbulent convection, the energy
transfer by heat conduction is negligible which means the Prandtl
number is very large. In all of our simulations, the $Pr$ is set to
be $10^4$.
 Therefore, approximately we have $\max{(Ra)}
\sim 6\times 10^8 \beta^2$. Similarly, we have Reynolds number
$Re=NM_a(1+M_a)\beta/\delta\sim 3\times10^2 \beta$. While in the
typical stellar convection, $Re\sim 10^{10}$. Hence,  $\beta\sim
10^8$ is needed when we do a direct numerical simulation (DNS) of
stellar convective flows where $Ra\sim 10^{24}$.  These values can
be reduced by increasing grids number which is very expensive for
the present generation hardware. At the same time, a very small
$\tau$ would introduce large computational error. An alternate way
is the LES which simulate the large eddies directly and approximate
the small eddies with models. There are a lot of approaches to
perform the LES, the simplest way may be the SGS model
(Smagorinsky~\cite{smagorinsky63}; Deardorff~\cite{deardorff71}).

In the BGK scheme, the viscosity is introduced through the
collisions of particles. The natural way of implementing SGS model
is to modify the collision time. Chen et al. (\cite{chen2003})
included the renormalization group $\tilde{k} - \epsilon$ large eddy
model into the BGK equation, where $\tilde{k}$ is the turbulent
kinetic energy and $\epsilon$ is the turbulent dissipation. In the
$\tilde{k} - \epsilon$ model, tow additional equations are needed to
be solved. For sake of simplicity, we consider the
Smagorinsky~(\cite{smagorinsky63}) model. In current study, the
collision time is defined as
\begin{equation}
\label{vis}
\tau_{tot}=\tau+C_2\frac{|p_l-p_r|}{|p_l+p_r|}\Delta t+\frac{\mu_{sgs}}{p},
\end{equation}
where the first term of right-hand side represents the molecule
viscosity and the second term is introduced to increase the
numerical dissipation when there is a jump in pressure around the
control volume boundary. $C_2$ is an adjustable constant, $p_l$ and
$p_r$ are the reconstructed pressures at the left and right side of
a cell interface (see sketch in Fig.~1). In the strong supersonic
region, additional dissipation caused by this term is essential to
stabilize the computing.

\begin{figure}[ht]
\centering
\label{split}
\includegraphics[width=8.0cm]{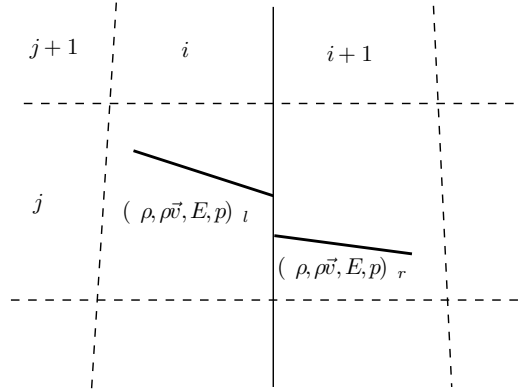}
\caption{Reconstruction of conservative variables and pressure where
discontinuity is introduced at the cell interface. The dashed lines
represent the boundaries of the of the control volumes numbered by
$i$ and $j$; the vertical solid line is the interface where the
flux-splitting is performed. The discontinuity of the values and
their slopes are depicted by oblique solid lines.}
\end{figure}

The last term in right-hand side of equation ({\ref{vis}}) is
implemented to account for the SGS viscosity,
\begin{equation}
\mu_{sgs}=\rho (c_{\mu}\Delta)^2(2\vec{\sigma}:\vec{\sigma})^{1/2},
\end{equation}
where $c_{\mu}$ is an adjustable constant, usually has the value:
0.1 $\sim$ 0.2, the filter width $\Delta$ is taken to be the local
resolution, colon stands for the contract of tensor and
$\vec{\sigma}=\partial_i v_j+\partial_j v_i$. Above model is called
Smagorinsky model or sometimes Smagoringsky-Lilly model.  In our
calculations, the eddy viscosity is computed in the control volume
by staggered mesh strategy and then interpolated to the cell
interface. In the BGK scheme, there is intrinsic diffusion caused by
particle collisions. In current study, we are only interested in the
turbulent properties. So, the molecule Prandtl number $Pr$ is set
very large. And the following diffusive flux (CS96) is implemented
explicitly into the BGK scheme,
\begin{equation}
\vec{F}_d=-C_T\nabla T -C_S \nabla S,
\end{equation}
where $S=C_p(\ln{T}-\nabla_a \ln{p})$ is the specific entropy,
$C_p$ is the specific heat at constant pressure
and $\nabla_a$ is the adiabatic gradient. In the stable layer, $C_T$ is set to
make the diffusion carry out the input energy flux. In the convection zone,
$C_T$ is very close to zero. $C_S=\mu_{sgs}/Pr_{sgs}$ represents the
turbulent diffusion and is set to zero in the stale region, where
$Pr_{sgs}$ is the effective Prandtl number of SGS turbulence and taken to be $1/3$.

\section{Physical Models}

\label{sect:model}
Our physical problem is very similar to those
studied by CS89 and CS96. An ideal gas ($p=\rho R T$)
 in a rectangular box is considered with gravity in the vertical
direction. The side boundaries are periodic and the top and bottom
boundaries are impenetrable and stress free. In order to avoid
boundary effects, a very thin stable layer is placed below the upper
boundary and the diffusive flux is gradually enhanced near the lower
boundary to make it carry out total flux at the lower boundary. A
constant energy flux $F_b=0.25$ is fed at the bottom. At the top,
the entropy is fixed. The system is initially static:
\begin{eqnarray}
T   &=& (1+\mathcal{Z}(d-z)/d)T_t,\\
\rho&=& (T/T_t)^n\rho_t,\\
p   &=& (T/T_t)^{n+1}p_t,\\
g   &=& (n+1)p_t\mathcal{Z}/(\rho_t d)
\end{eqnarray}
where $\mathcal{Z}=(T_b-T_t)/T_t$ is the normalized parameter, $0\le
z\le 1$ and $n$ is the polytropic gas index.  The gravitational
acceleration $g$ comes from the hydrostatic equilibrium $\partial
p/\partial z=-\rho g$. The subscripts $t$ and $b$ denote top and
bottom values respectively. Above solutions to Navier-Stokes
equations are not stable against small perturbations. In all of our
calculations, the velocity field is slightly perturbed initially.
After long-time thermodynamic relaxation, the system will reach a
statistical steady state. We defined a series of runs to test the
effects of numerical parameters. The numerical effects of a variety
of parameters were studied by Chan \& Sofia~(\cite{chan86},
hereafter CS86)
 and CS89 in details. The
effects of changing turbulent Prandtl number was tested by Singh \&
Chan(\cite{singh93}). Here, we just focus on the important parameter
$c_\mu$ in the SGS model and new parameter $C_2$ appearing in the
BGK scheme. The details are given in the second line of
Table~{\ref{Tab:results}}.

All the cases we computed use a $29\times 29 \times 45$ mesh. The
vertical grid decreases smoothly with height (about $6$ grids per
PSH) and
 the horizontal grid is uniform. The aspect ratio (width/depth) of the box is $1.5$.

\section{Results}

\label{sect:analysis}
\begin{figure}[ht]
\centering
\label{flxs}
\includegraphics[width=8cm]{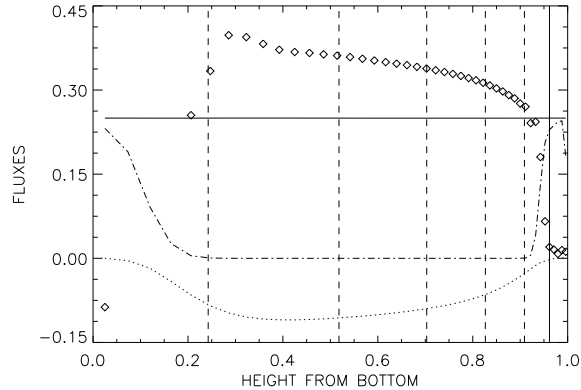}
\caption{Distribution of temporally and horizontally averaged energy
fluxes with depth. Dash dot line: diffusive flux; dotted line:
kinetic energy flux; diamonds: enthalpy flux; solid line: total
energy flux.}
\end{figure}
In this part, we show the results from the numerical simulations.
All the runs were evolved $2000000$ numerical time-steps,
corresponding to a dimensionless time around $858$, before the
statistical analysis is performed. The statistical steady state is
indicated by the balance of the input energy flux from the bottom
and the outgoing energy flux through the top. In our calculations,
the spatial variation of averaged total energy flux from $0.25$ is
within $0.1\%$ (see solid line in Fig.~2). Another criterion is the
averaged vertical mass flux which is less then $10^{-5}$ everywhere
in all cases. Hence, the system will not undergo substantial
adjustment any more. The statistical average covers $500000$
numerical time-steps.

Except for the instant velocity fields, all the other quantities
investigated here are the mean values. For an arbitrary quantity
$q$, $\al q \ar$ represents its combined horizontal and temporal
mean, $q'$ denotes the deviation from  $\al q \ar$, $q''$ stands for
the root mean square (rms) fluctuation from the $\al q \ar$.

In some figures, the integral pressure scale height (PSH) is shown
by the vertical lines. For example, in Fig.~2, the vertical solid
line denotes the location of stable-unstable interface near the
upper boundary. The second dashed line at the left side of the solid
line is 2 PSHs away from the upper stable-unstable interface.
Although the numerical parameters are different for case A to case
C, their relaxed thermal structure are nearly equivalent and the
discrepancy between the integral PSH locations is very small. So, we
can plot the results from all runs in the same figures.

\subsection{Velocity Fields}
\label{sect:velfie}

\begin{figure}[ht]
\centering
\label{vfs}
\includegraphics[width=7.2cm]{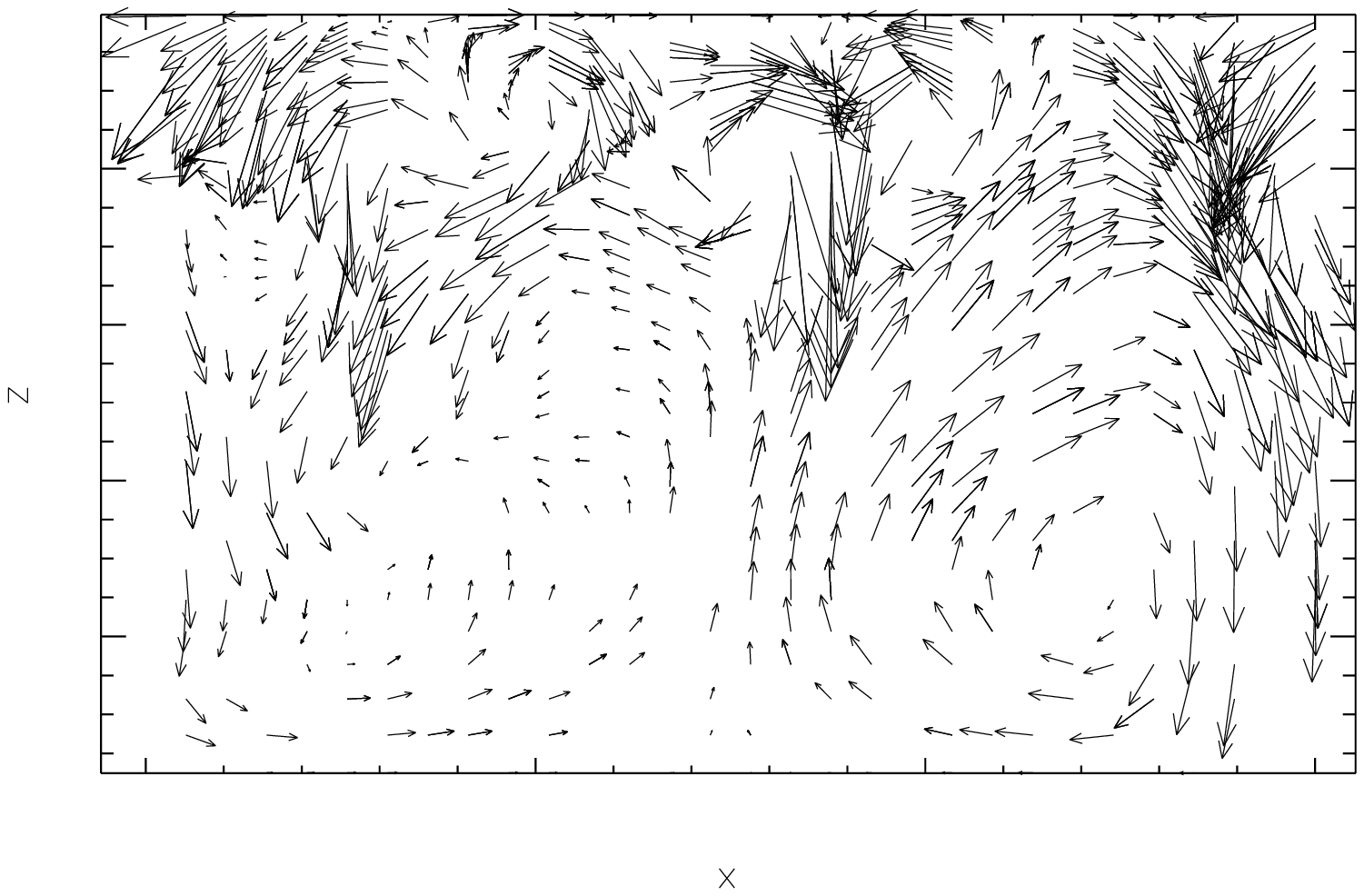}
\includegraphics[width=7.2cm]{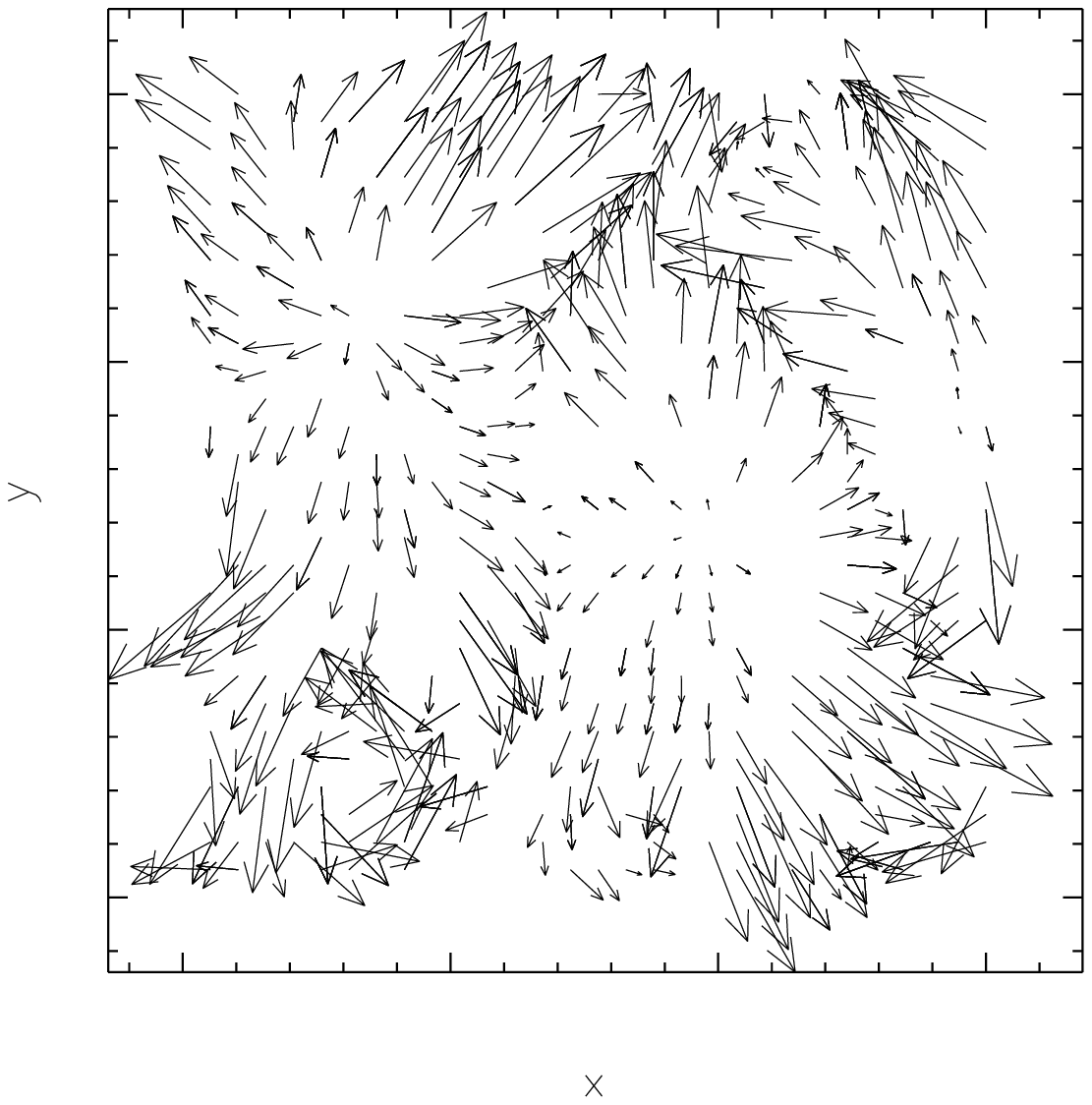}
\caption{Instant velocity fields at time $858$.  Left: projection in
vertical plane at $y=0.45$; right: projection in horizontal plane at
$z=0.6$. }
\end{figure}
In CS86, the three-dimensional turbulent flow structure
 was well depicted by the pseudo stream lines.
For sake of clarity, Fig.~3 shows the velocity fields projected in
x-z plane and x-y plane. From the left panel of Fig.~ 3, it is
evident to see that the high speed motions exist in the top region
and are associated with the downward streams. In our calculations,
the maximum Mach number $Ma$ occasionally exceeds one. In CS96, the
SGS viscosity was enhanced by a factor
$[1+0.5(v_x^6+v_y^6+v_z^6)/c_s^6]$ to suppress the shocks occurring
in the top region which would easily trigger the instability of the
numerical computation, especially, during the early thermal
relaxation. Based on flux-splitting (see Fig.~1), such implement can
be avoided in gas-kinetic BGK scheme. If the nonlinear van Leer
limiter is replaced by central interpolation, the supersonic motion
cannot be handled correctly.  Numerical tests also show that a $Ra$
less than $10^{9}$ will smooth the turbulence, the circulations
become laminar. The solutions are almost unchanged when $Ra$ great
than $10^{15}$. In current study, we adopt $Ra=10^{19}$. The
networks of downward streams can be seen the right panel of Fig.~3.

\subsection{Approximate Relations}

\begin{sidewaystable*}[htp]
  \caption[]{Quantitative Estimates of Some Approximate Relations.}
  \label{Tab:results}
  \begin{center}\begin{tabular}{lrrrrrll}
  \hline\noalign{\smallskip}
Relation\qquad\quad Parameter &   A\ \ \ \ \      & B\ \ \ \ \    & C\ \ \ \ \    &  CS89\ \ \ \ \ & Deviation & ID & Category \\
  \hline\noalign{\smallskip}
 \qquad\qquad\qquad\qquad $c_\mu$ & 0.2\ \ \ \ \  & 0.25\ \ \ \ \ & 0.2\ \ \ \ \  &  0.2\ \ \ \ \  &(\%) & &  \\
 \qquad\qquad\qquad\qquad  $C_2$  &   1\ \ \ \ \  & 1\ \ \ \ \    & 0\ \ \ \ \    &  - - -\ \ \ \ \ & & & \\
  \hline\noalign{\smallskip}
$v''_x/v''_z$                         &  0.69 ( 0.12 ) & 0.66 ( 0.11 ) & 0.62 ( 0.11 ) & 0.61 ( 0.05 ) &   8& R0$^\dag$  &III\\
$v''_y/v''_z$                         &  0.57 ( 0.10 ) & 0.58 ( 0.10 ) & 0.64 ( 0.11 ) & 0.61 ( 0.05 ) &   2& R1  &III\\
$(\rho''/\al\rho\ar)/(T''/\al T\ar)$  &  0.83 ( 0.01 ) & 0.83 ( 0.02 ) & 0.83 ( 0.01 ) & 0.89 ( 0.04 ) &   7& R2  &I  \\
$(p''/\al p\ar)/(T''/\al T\ar)$       &  0.57 ( 0.04 ) & 0.56 ( 0.04 ) & 0.57 ( 0.04 ) & 0.57 ( 0.07 ) &   1& R3  &I  \\
$S''/(C_pT''/\al T\ar)$               &  0.90 ( 0.01 ) & 0.90 ( 0.01 ) & 0.90 ( 0.01 ) & 0.94 ( 0.03 ) &   4& R4  &I  \\
$(p''/\al p\ar)/(v''^2/\al T\ar)$     &  0.48 ( 0.01 ) & 0.47 ( 0.01 ) & 0.48 ( 0.00 ) & 0.26 ( 0.01 ) &  83& R5$^\ddag$ &II \\
$p''/\al\rho\ar v''^2_z$              &  0.88 ( 0.14 ) & 0.85 ( 0.13 ) & 0.87 ( 0.14 ) & 0.51 ( 0.03 ) &  70& R6  &IV \\
$T''/v''^2_z$                         &  1.52 ( 0.15 ) & 1.51 ( 0.13 ) & 1.51 ( 0.15 ) & 0.90 ( 0.10 ) &  68& R7  &IV \\
$C(T',S')$                            &  0.98 ( 0.00 ) & 0.98 ( 0.02 ) & 0.98 ( 0.03 ) & 0.99 ( 0.01 ) &   1& R8  &I  \\
$C(\rho',S')$                         & -0.93 ( 0.01 ) &-0.92 ( 0.01 ) &-0.92 ( 0.01 ) &-0.89 ( 0.06 ) &   4& R9  &I  \\
$C(\rho',T')$                         & -0.83 ( 0.02 ) &-0.83 ( 0.02 ) &-0.82 ( 0.03 ) &-0.82 ( 0.05 ) &   1& R10 &I  \\
$C(p',T')$                            &  0.59 ( 0.04 ) & 0.59 ( 0.04 ) & 0.59 ( 0.03 ) & 0.49 ( 0.05 ) &  20& R11 &II \\
$C(v_z,T')$                           &  0.78 ( 0.02 ) & 0.79 ( 0.02 ) & 0.77 ( 0.02 ) & 0.81 ( 0.03 ) &   4& R12 &I  \\
$C(v_z,S')$                           &  0.76 ( 0.02 ) & 0.78 ( 0.03 ) & 0.75 ( 0.02 ) & 0.81 ( 0.03 ) &   6& R13 &I  \\
$C(v_z,\rho')$                        & -0.64 ( 0.01 ) &-0.67 ( 0.01 ) &-0.63 ( 0.01 ) &-0.74 ( 0.03 ) &  13& R14 &II \\
$\al v_z\rho'\ar/\al v_z\ar\al\rho\ar$& -1.00 ( 0.02 ) &-0.99 ( 0.02 ) &-1.00 ( 0.02 ) &-1.00 ( - - - ) &   0& R15 &I  \\
$\al v_z p\ar/\al v_z\ar\al p\ar$     &  1.49 ( 0.06 ) & 1.44 ( 0.05 ) & 1.50 ( 0.06 ) & 1.24 ( 0.08 ) &  19& R16 &II \\
$\al v_zT'\ar/\al v_z\ar\al T\ar$     &  1.46 ( 0.06 ) & 1.41 ( 0.06 ) & 1.46 ( 0.06 ) & 1.26 ( 0.08 ) &  15& R17 &II \\
$\al v_zS'\ar/C_p\al v_z\ar$          &  1.28 ( 0.05 ) & 1.24 ( 0.04 ) & 1.28 ( 0.05 ) & 1.20 ( 0.08 ) &   6& R18 &I  \\
$\al v_z\ar/\al {v''}_z^3\ar\al T\ar$ &  0.81 ( 0.08 ) & 0.85 ( 0.07 ) & 0.80 ( 0.09 ) & 0.58 ( 0.07 ) &  41& R19 &IV \\
$F_{ep}/C_p\al p\ar\al v_z\ar$        &  1.49 ( 0.06 ) & 1.44 ( 0.05 ) & 1.50 ( 0.06 ) & 1.25 ( 0.08 ) &  18& R20 &II \\
$F_{ep}/C_p\al\rho\ar{v''}_z^3$       &  1.20 ( 0.15 ) & 1.23 ( 0.13 ) & 1.20 ( 0.15 ) & 0.72 ( 0.06 ) &  68& R21 &IV \\
$T''/\al T\ar=a \Delta\nabla+b$ &  $a=1.92$   & $a=1.88$ &   $a=1.95$   & $a=1.05$                     &  83& R22 &II \\
                                &  $b=0.0029$ & $b=0.0025$ & $b=0.0030$ & $b=0.0027$ &   &  &\\
                                &  ( $0.0004^\ast$ ) & ( $0.0001^\ast$ ) & ( $0.0002^\ast$ ) & ( $0.0008^\ast$ ) &  &  &\\
$v''/\al T\ar=a \Delta\nabla+b$ &  $a=1.01$   & $a=1.03$ &   $a=1.03$   & $a=1.17$                     &  13 & R23 &II\\
                                &  $b=0.0027$ & $b=0.0023$ & $b=0.0028$ & $b=0.0032$ &   &  &\\
                                &  ( $0.0004^\ast$ ) & ( $0.0002^\ast$ ) & ( $0.0004^\ast$ ) & ( $0.0008^\ast$ ) &  &  &\\
$\Delta\nabla=a [F_b/(0.8C_p\al p\ar\al T\ar^{1/2})]^{2/3}+b$
                                & $a=0.71$   & $a=0.71$    & $a=0.70$    & $a=0.9$                     &  21 & R24 &II\\
                                & $b=-0.0013$ & $b=-0.0011$ & $b=-0.0013$  & $b=-0.002$ &   &  &\\
                                &  ( $0.0002^\ast$ ) & ( $0.0002^\ast$ ) & ( $0.0001^\ast$ ) & ( $0.0008^\ast$ ) &  &  &\\
  \noalign{\smallskip}\hline
  \end{tabular}
\begin{list}{}{}
\item[$^\dag$]  New relation defined in current paper.
\item[$^\ddag$] ${v''}^2={v''}_x^2+{v''}_y^2+{v''}_z^2$.
\item[$^\ast$]  These are standard deviations of the least-squares fits.
\end{list}
\end{center}
\end{sidewaystable*}
Our BGK code has been extensively tested for laminar flows (Tian et
al.~\cite{tian2007}). In order to validate the incorporation of SGS
model, we re-estimate quantitatively
 some approximate relations among the thermodynamic variables and their fluctuations. The results
are given in Table~{\ref{Tab:results}} where the results from CS89
are also listed for comparison. Our computational models are
partially similar to CS89 and partially to CS96. Models of CS89
undergo substantially adjustment near the boundaries and we cannot
afford the high-resolution desired in CS96.  During the data
analysis, the correlation function of quantity $q$ and $p$ is
defined as $C[q,p]=\al qp\ar/({\al q^2\ar^{1/2}}{\al
p^2\ar^{1/2}})$. The standard deviations of these approximations
($\sigma_1$) are given in the brackets. The deviation of $Ri$ from
CS89 is calculated by
$\sigma_2=|((Ri_A+Ri_B+Ri_C)/3-Ri_{CS89})/Ri_{CS89}|$. We only
concentrate on the middle region in the convection zone, namely, 1
PSH from the bottom and 2 PSHs from the upper stable-unstable
interface. The investigated layer expands about 3 PSHs.
\begin{figure}[ht]
\centering
\label{rms}
\includegraphics[width=7.2cm]{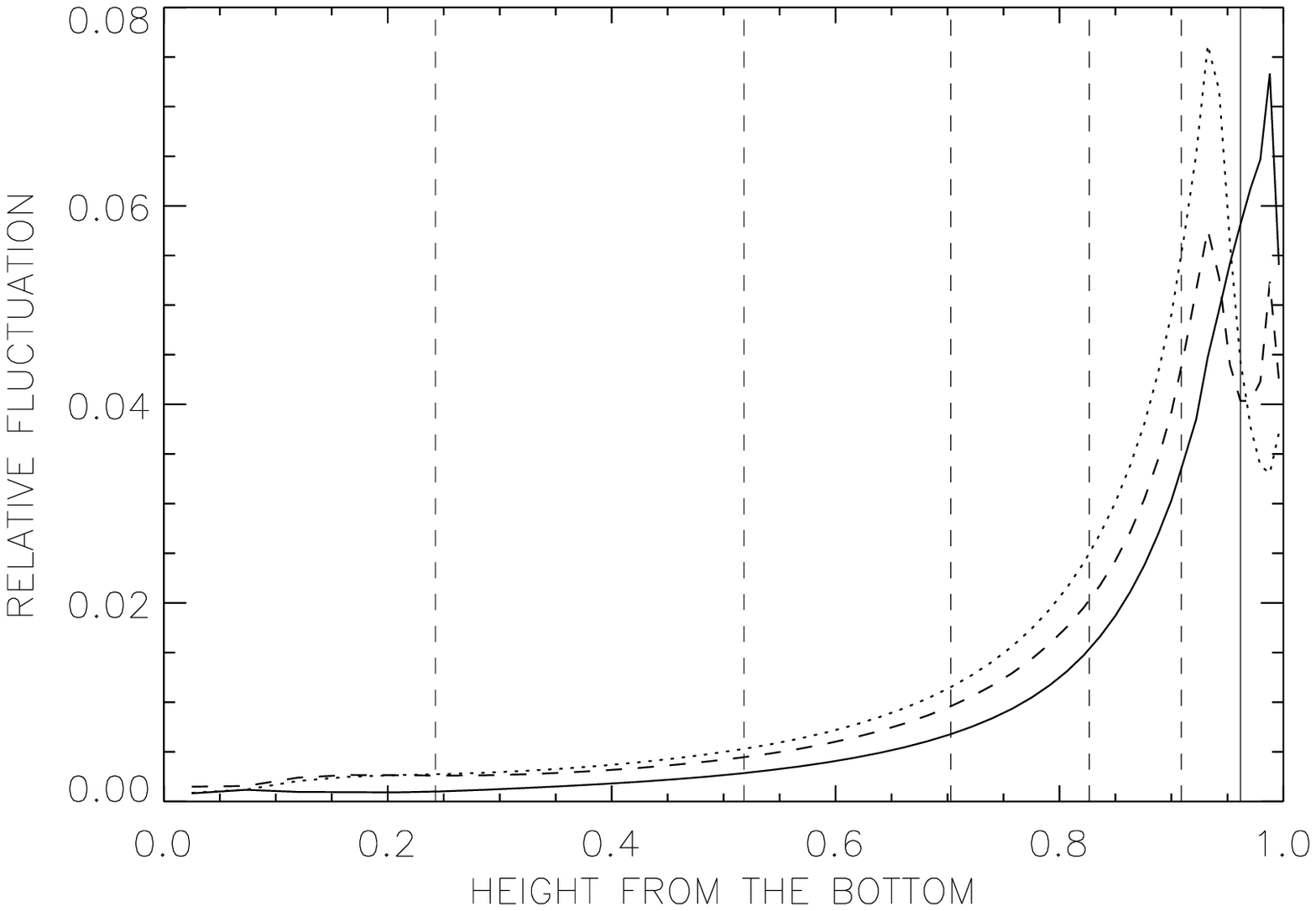}
\includegraphics[width=7.2cm]{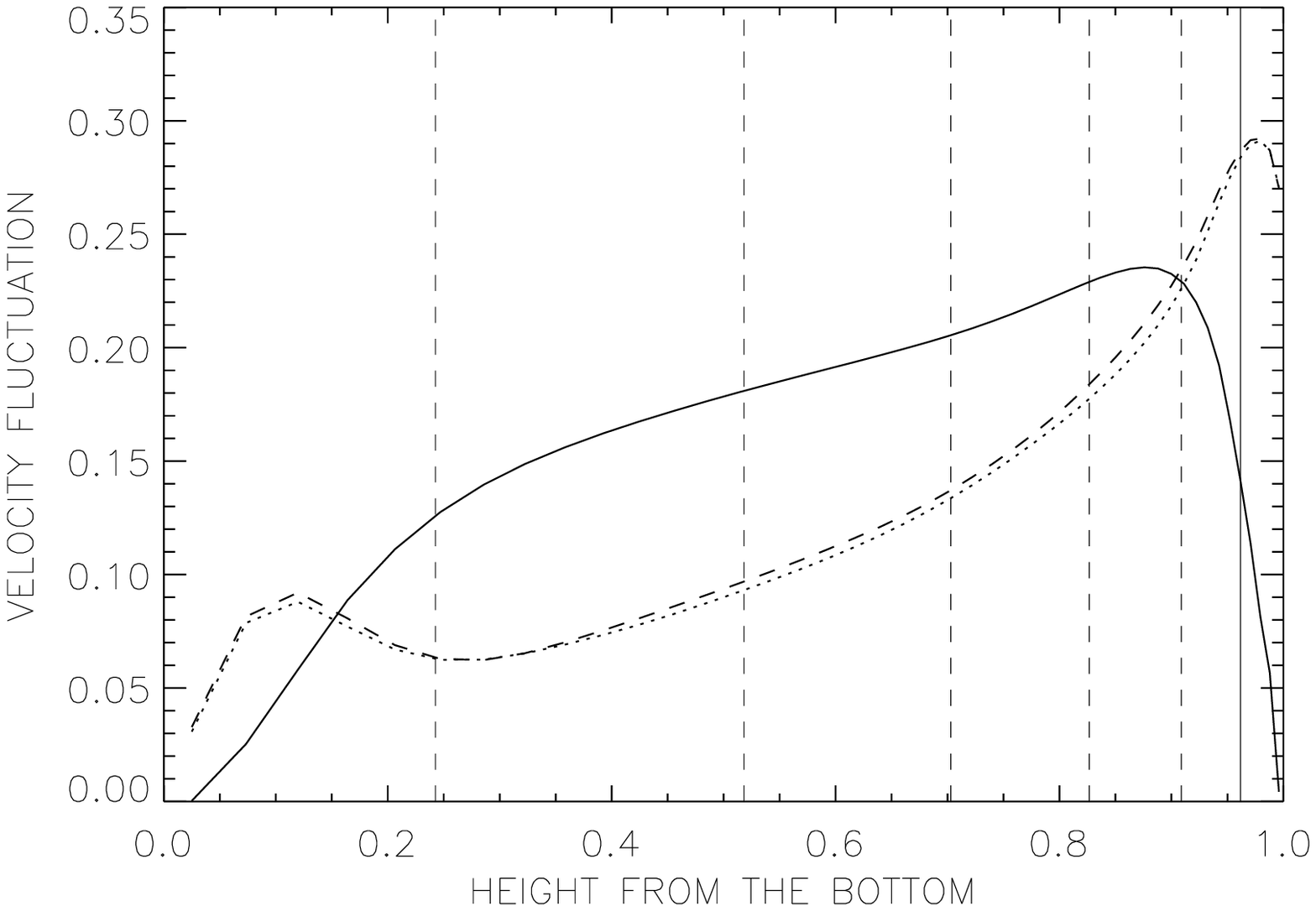}
\caption{Height distributions of fluctuations and relative
fluctuations for case C. Left: $p''/\al{p}\ar$ (solid line),
$T''/\al{T}\ar$ (dotted line), $\rho''/\al{\rho}\ar$ (dashed line);
right: $v''_x$ (dashed line), $v''_y$ (dotted line), $v''_z$ (solid
line).}
\end{figure}
The approximate relations and correlations in
Table~\ref{Tab:results} are classified roughly into four categories
which are indicated by Roman numerals in the last column according
to the goodness of fit and discrepancy from CS89. These categories
are: Category I:   $\sigma_1\le 0.6$ and $\sigma_2\le 10$\%;
Category II:  $\sigma_1\le 0.6$ and $\sigma_2 >  10$\%; Category
III: $\sigma_1 >  0.6$ and $\sigma_2\le 10$\%; Category IV:
$\sigma_1 >  0.6$ and $\sigma_2 >  10$\%.

Category I contains the best relations where the thermal variables
are mainly involved, i.e. $\rho$, $p$, $T$ and their fluctuations.
These relations should be weekly affected by numerical scheme and
details of the computational models.  In the work of Kim et al.
(\cite{kim95}) where a realistic equation of state (EOS) was used,
these relations are different from current studies. Therefore,
Category I may  be dominantly determined by EOS.

Most relations of Category II are functions of pressure $p$ ,
velocity $\vec{v}$ and their fluctuations. The amplitude of the
fluctuations of the components of velocity in our computation is
lower than those in CS89 (Fig.~1 therein) and CS96 (Fig.~2 therein)
while the amplitude of the relative fluctuations of thermal
quantities are around two times larger than those in CS89 (see
Fig.~2 therein) and in Singh's work (\cite{singh93}).  Hence when
the fluctuations of thermal variables are expressed in terms of
$v''$, $v''_z$, etc., the coefficients are nearly doubled, e.g.,
$R5$ and $R22$. This can also be seen in Fig.~3a in CS96.

Ratios of $v''_x/v''_z$ and $v''_y/v''_z$ do not deviate from CS89
very much, but the fitting is not accurate. This may be caused by
the effects of the boundary and transition layers. It is obvious
from Fig.~4 that the height distribution of $v''_z$ is not flat as
CS89. There is a small hump below the upper transition layer (one
PHS from the unstable-stable interface) which can also be found in
CS96. We believe that the unstable-stable transition  is responsible
for this. The large difference in Category IV should be caused by
the high order powers of velocity fluctuations.

$R22$ $\sim$ $R24$ are commonly approximated in MLT. In current
study, these relations are fitted by linear approximation very well.
However, the slopes are different from the CS86 and CS96. The
situation of $R22$ can be explained as above. For $R23$,
 the reasons may lie in
the fact that in the upper convective region, the amplitude of
$v''_z$ from CS96 is larger than ours and therefor we need a smaller
$R23$. CS96 got a smaller $R24$ (0.78) than CS89. In current study,
it is even smaller which implies a smaller super-adiabatic gradient
$\Delta\nabla$.

The cause of most of the discrepancies between current study and
CS89 may be  the larger $T''$ and smaller $v_z''$. In the
traditional theory of stellar convection, the enthalpy flux is
proportional to the fluctuations of temperature and velocity, i.e.,
$F_e \sim C_p v'' T''$ and the kinetic flux is totally neglected.
When the kinetic flux is comparable to the total flux, this kind of
proportional relation is not exactly held any more. Based on the
following facts:
\begin{enumerate}
  \item $F_b=0.0625$, $\max{T''}\sim 0.022$, $\max{v''_z}\sim 0.2$
(Singh~\cite{singh93}, Fig.~1 and Fig.~5);
  \item $F_b=0.1250$,
$\max{T''}\sim 0.023$, $\max{v''_z}\sim 0.22$ (CS89, Fig.~1 and
Fig.~2);
  \item $F_b=0.2500$, $\max{T''}\sim 0.065$, $\max{v''_z}\sim 0.35$
(CS96, Fig.~2);
  \item $F_b=0.2500$, $\max{T''}\sim 0.078$, $\max{v''_z}\sim
0.23$ (current, Fig.~4);
  \item $F_b=0.6551$, $\max{T''}\sim 0.129$,
$\max{v''_z}\sim 0.27$ (Kim~\cite{kim95}, Fig.~7 and Fig.~8),
\end{enumerate}
we can conclude that the $T''$ and $v_z''$ are proportional to $F_b$
in a nonlinear way. Note that current study use a different
numerical scheme and Kim adopted a realistic EOS. It is beyond the
scope of current study to perform a quantitative analysis of such
relations.

 Above comparison shows that the dynamical properties and thermal
properties for stellar type of convection may be affected by
different aspects of the numerical models separately. Thermal
structure is mainly determined by physical parameters while dynamic
motions can easily affected by numerical parameters.

\begin{figure}[ht]
\centering
\label{xvz}
\includegraphics[width=8cm]{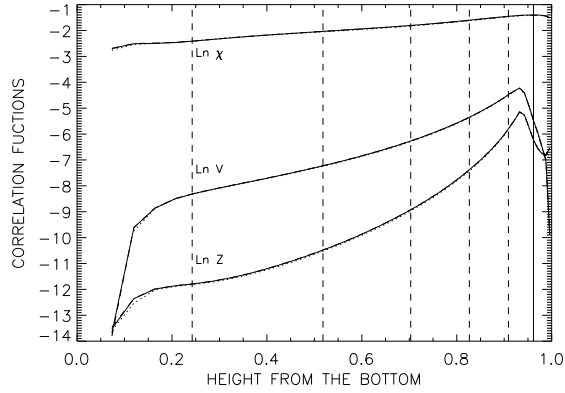}
\caption{Height distributions of natural logarithm of $\chi$, $V$ and $Z$.
Solid line: case A; dotted line: case B; dashed line: case C.}
\end{figure}

\subsection{Anisotropic Turbulence}

Xiong's non-local time-dependent stellar convection theory is based
on Reynolds stress method. It is dynamic theory of auto- and
cross-correlation functions of turbulent velocity and temperature
fluctuations. These fluctuations are defined as the derivation from
the density weighted average, i.e.,
\begin{equation}
u'_i=v_i-\frac{\al \rho v_i\ar}{\al\rho\ar},\quad
\tilde{T}'=T-\frac{\al\rho T\ar}{\al\rho\ar}.
\end{equation}
The start point of Xiong's theory is a set of partial differential
equations of
\begin{equation}
\chi^2=\al{w'_iw'^i}\ar/3,\quad
Z=\al{\tilde{T}'^2}\ar/\al{\tilde{T}}\ar^2,\quad
V=\al{\tilde{T}'w'_i}\ar/\al{\tilde{T}}\ar,
\end{equation}
where $w'_i=\rho u'_i/\al{\rho}\ar$, $\al{\tilde{T}}\ar={\al\rho
T\ar}/{\al\rho\ar}$ and the summation convention is adopted. The
numerical results of $\chi$, $Z$ and $V$ are given in Fig.~5. In the
closure models of Xiong's theory, three adjustable parameters, i.e.,
$c_1$, $c_2$ and $c_3$ are introduced. $c_3$ is used to describe the
anisotropic turbulent motions. In Deng's (\cite{deng2006}) work,
$c_3$ was related to turbulent velocity by
${w'}_z^2/({w'}_x^2+{w'}_y^2)=(3+c_3)/2c_3$ in the fully unstable
zone. In the upper overshooting region, they proposed that
${w'}_z^2/({w'}_x^2+{w'}_y^2)\sim 0.5$ and is independent of $c_3$.
In the lower overshooting zone, ${w'}_z^2/({w'}_x^2+{w'}_y^2)\le
0.5$ and decrease as $c_3$ decrease.  There is no enough room for
overshooting in our models. The ratio of
${w'}_z^2/({w'}_x^2+{w'}_y^2)$ from numerical simulations is given
in Fig.~6 from which we can see this ratio is slightly dependent on
$c_\mu$ and $C_2$. In the upper efficient-inefficient convection
interface (about 1 PSH from unstable-stable interface), this value
approximately equals to $0.5$. It takes its maximum at the location
where the turbulent convection starts to become inefficient near the
bottom. Its maximum is affected evidently by $c_\mu$ and $C_2$.
Current study cannot give a definite solution for anisotropic
turbulence. Here, we present preliminary suggestions. Suppose that
${w'}_z^2/({w'}_x^2+{w'}_y^2)=(3+c_3)/2c_3$ is held in current
models, we can conclude that $c_3$ is infinitely large at the upper
efficient-inefficient interface and decreases as the distance from
the boundaries increases. The minimum of $c_3$ is about $0.75$.
\begin{figure}[ht]
\centering \label{anv}
\includegraphics[width=8cm]{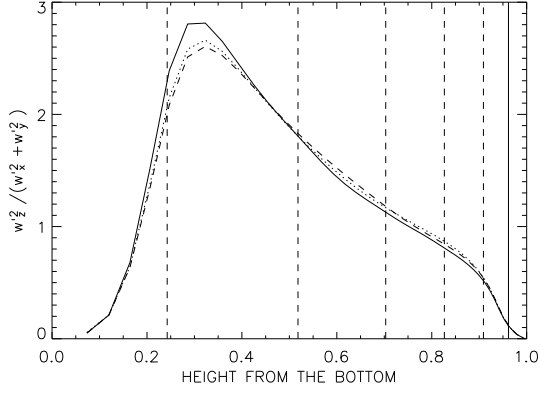}
\caption{Height distribution of ${w'}_z^2/({w'}_x^2+{w'}_y^2)$.
Solid line: case A; dotted line: case B; dashed line: case C.}
\end{figure}
\subsection{Non-local Transport Models}

\begin{figure}[ht]
\centering
\label{ntes}
\includegraphics[width=7.2cm]{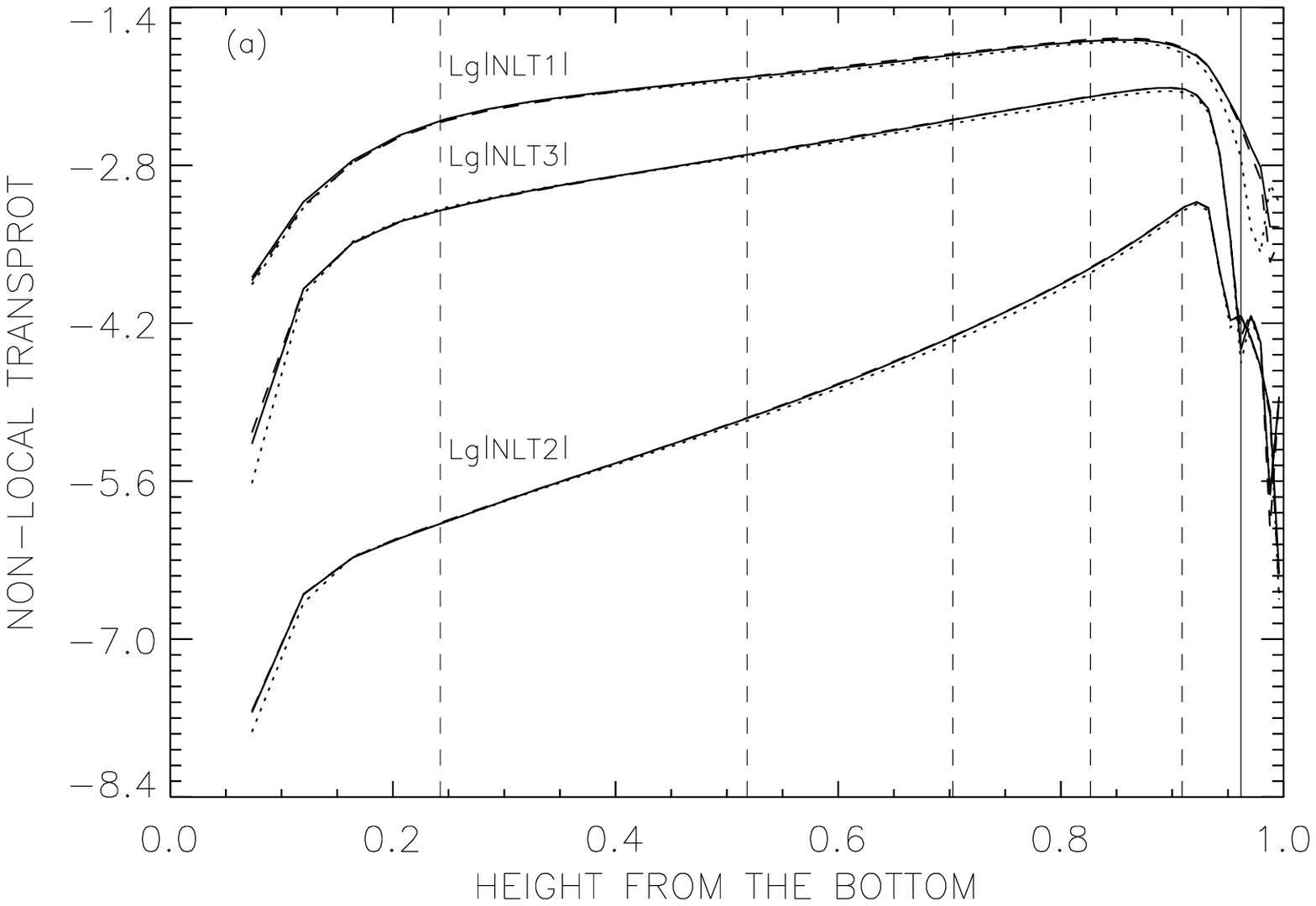}
\includegraphics[width=7.2cm]{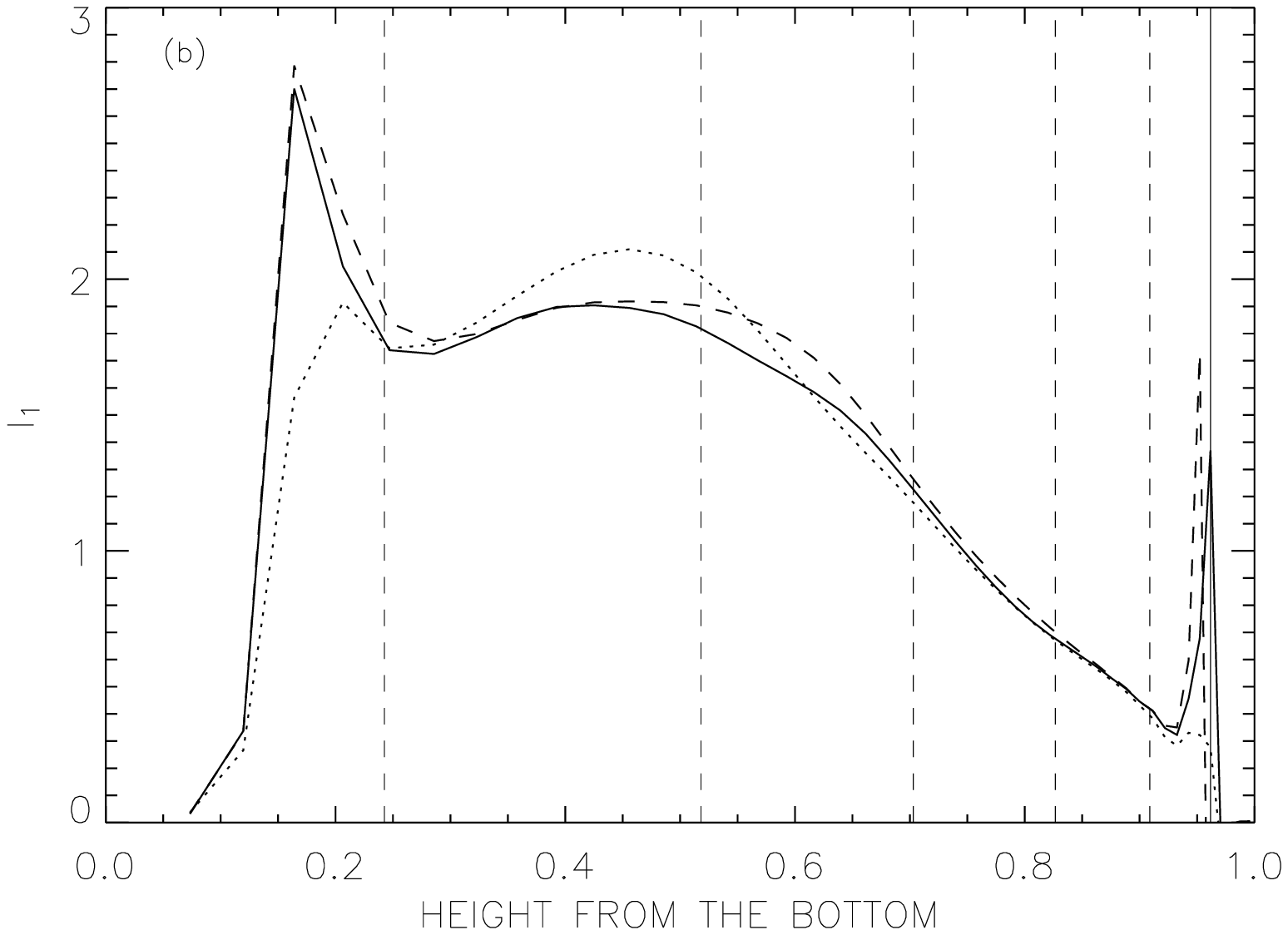}
\includegraphics[width=7.2cm]{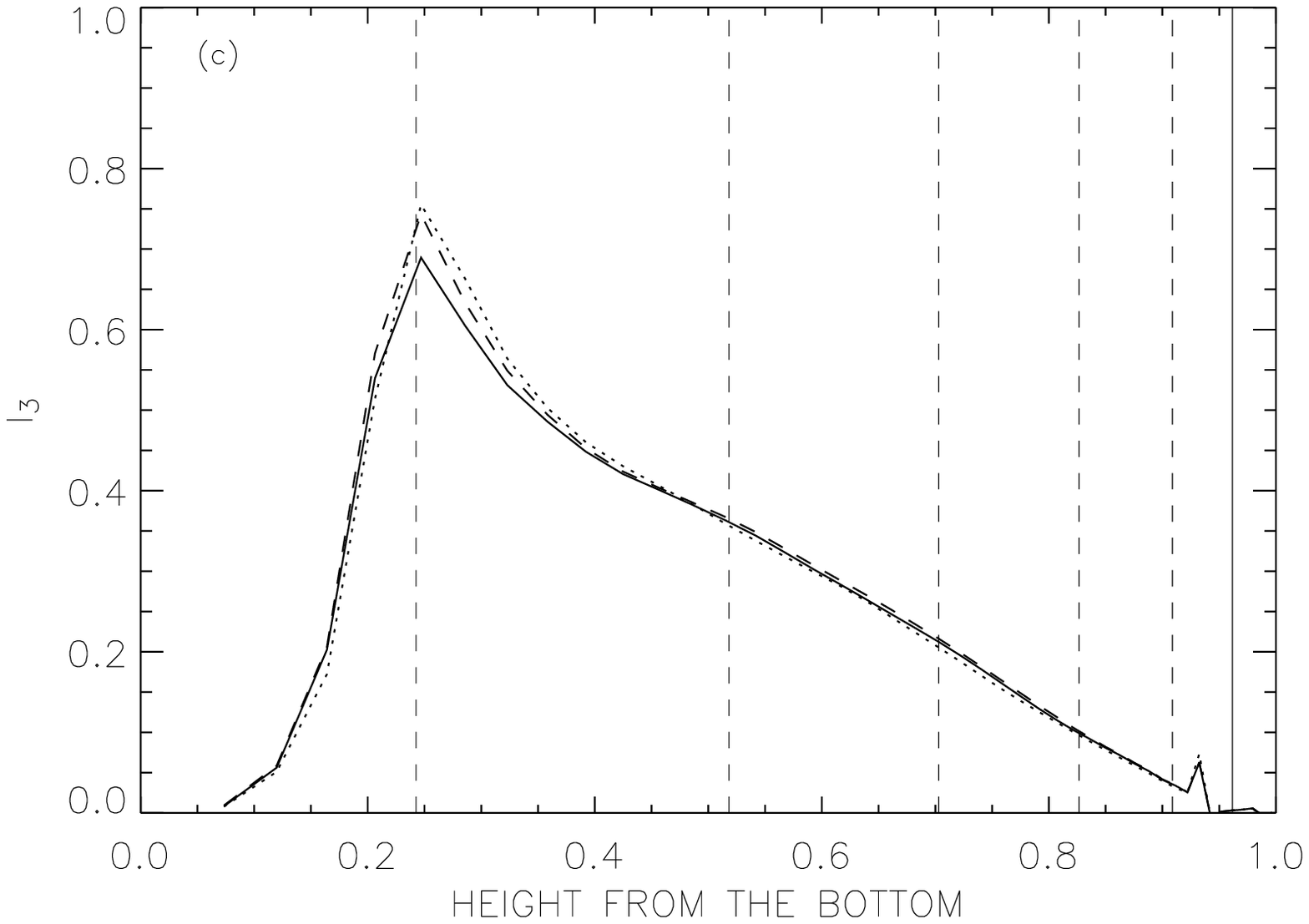}
\includegraphics[width=7.2cm]{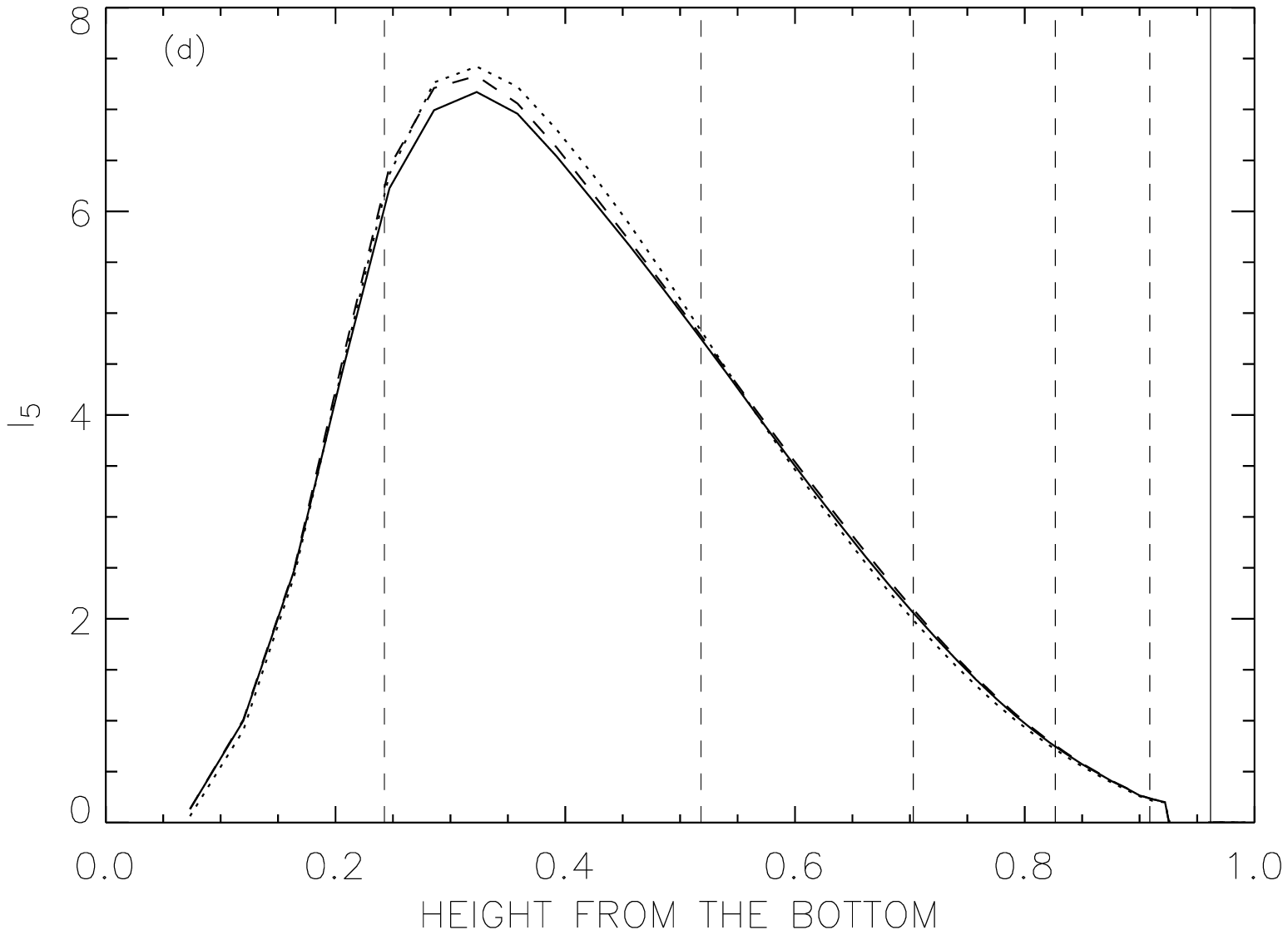}
\caption{(a) Non-local turbulent transports (in common logarithm).
(b)$\sim$(d) The approximate coefficients. Solid line: case A;
dotted line: case B; dashed line: case C.}
\end{figure}
Generally, a Reynolds stress method suffers the so-called closure problem.
 CS96 numerically studied
the popular closures and found they were poor. Some third order
moments representing the non-local transport effects were
approximated by gradient models in Xiong's work (\cite{xiong89}),
i.e.,
\begin{eqnarray}
NLT1&=&\al{u'_k w'_i w'^i}\ar=-\chi l_1 \nabla_k \al{w'_i w'^i}\ar,\\
NLT2&=&\al{u'_k \tilde{T}'^2}\ar/\al{\tilde{T}}\ar^2
=-\chi l_3 \nabla_k (\al{\tilde{T}'^2}\ar/\al{\tilde{T}}\ar^2),\\
NLT3&=&\al{u'_k w'^i \tilde{T}'}\ar/\al{\tilde{T}}\ar=-\chi l_5
\nabla_k (\al{w'^i\tilde{T}'}\ar/\al{\tilde{T}}\ar),
\end{eqnarray}
with $l_1\simeq l_3 \simeq l_5=\Lambda$, where $\Lambda$ is the
Lagrangian integral length scale of turbulence. The non-local
transports and coefficients from numerical simulations for all cases
are shown in Fig.~7 . From panel (a) of Fig.~7, we can see that the
$|\al{u'_k w'_i w'^i}\ar|$ is about one order larger than $|\al{u'_k
w'^i \tilde{T}'}\ar/\al{T}\ar|$  and $|\al{u'_k
\tilde{T}'^2}\ar/\al{T}\ar^2|$ is three orders less than $|\al{u'_k
w'^i \tilde{T}'}\ar/\al{T}\ar|$. Hence, the non-local transports are
dominated by turbulent kinetic energy.  Panel (b)$\sim$(d)  of
Fig.~7 show that the $l_1$, $l_3$ and $l_5$ gradually increase with
the distance from the top boundary and change  this trend near the
interface where  the turbulent convection become inefficient. The
variation of $l_1$ is slow in the lower half convection zone around
the value of 1.8. But it is affected by $c_\mu$ and $C_2$ obviously.
$l_5$ is about ten times larger than $l_3$. Both of $l_3$ and $l_5$
vary rapidly which suggests that there may not exist the universal
constant for these closure models.

\subsection{Effects of Numerical Parameters: $c_\mu$ and $C_2$}

\begin{figure}[ht]
\centering
\label{ars}
\includegraphics[width=8cm]{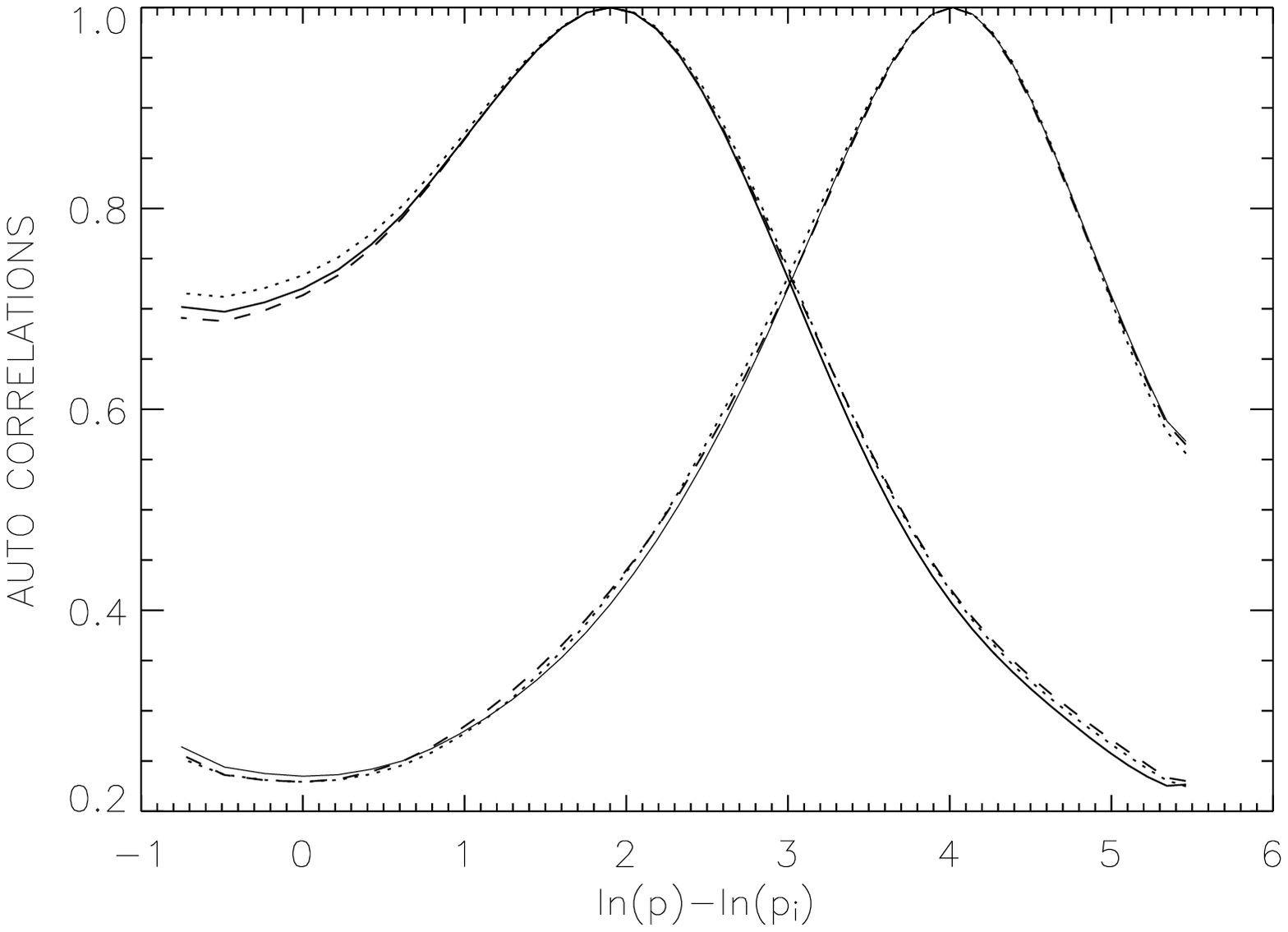}
\caption{Comparison of auto-correlations of the vertical velocity.
Solid line: case A; dotted line: case B; dashed line: case C. $p_i$
is the pressure at the unstable-stable interface.}
\end{figure}

In current LES, the local effective grid Reynolds number is enlarged
by SGS model whose amplitude is controlled by Deardorff constant
$c_\mu$. An inadequately small $c_\mu$ would cause the building-up
of the kinetic energy at the two-grid level and make the computation
crashed. If $c_\mu$ is too large, the turbulent motions will be
damped down.  The proper value of $c_\mu$ is resolution- and
method-dependent since the numerical dissipation also plays an
important role in the behavior of turbulence. Deardorff
(\cite{deardorff71}) suggested a value of $0.2$ for $c_\mu$ which
was used in the CS89. In CS96, this value was increased to $0.25$.

In the flux-splitting method, the discontinuity is introduced at the
cell interface (see Fig.~1) by limiters. Additional dissipation (the
second term in the left hand side of equation (\ref{vis})) is
employed to handle the strong shock waves near the sharp jumps of
pressure. The typical value of $C_2$ is 1. In the smooth region this
kind of discontinuity should be very small. However, our simulations
have lower resolution and the computing zone extends about seven
PSHs. So it is necessary to check if these values are adequate for
studying the stellar type of convection.

Relations $R2$ $\sim$ $R4$ and $R9$ $\sim$ $R11$ in
Table~\ref{Tab:results} show that the effects of changing these
parameters on the thermal structure are really slight. They both
affect the eddy properties with very small amplitude which can be
seen from Fig.~8 where the auto-correlations of the vertical
velocity are shown. Their profiles are nearly symmetrical except in
the upper stable and transition zone. The half width at half maximum
(HWHM) of these profiles should be sensitive on the viscosity
(CS86). $R0$ and $R1$ in Table~\ref{Tab:results} are the
$v''_x/v''_z$ and $v''_y/v''_z$, respectively. They should be nearly
equal to each other for isotropic turbulent flows. In our study, for
case A ($c_\mu=0.2$, $C_2=1$), the discrepancy between them is clear
which becomes slight for case B ($c_\mu=0.25$, $C_2=1$) and further
slight for case C ($c_\mu=0.2$, $C_2=0$). This kind of anisotropy
comes from the initially perturbation and can also bee found in
Fig.7 and Fig.11 in CS86. So it seems that $c_\mu=0.25$ is more
suitable for current study. In the deep stellar convection, the
shock waves are mild. Hence, the flux-splitting may enough to handle
it as in our tests. We suggest that it is better to make $C_2$ as
small as possible because it would enhance the anisotropy which can
be diminished by larger $c_\mu$.

\section{Conclusions}
\label{sect:conclusion}

In this paper, we present a preliminary application of gas-kinetic
BGK scheme to the simulation of turbulent convection in stellar
atmosphere. The approximate relations among thermodynamic variables,
their fluctuations and correlations were examined. The anisotropy
and diffusive models of non-local transport were investigated too.
The effects of varying numerical parameters were also tested. The
main conclusions are summarized as follows:
\begin{enumerate}
  \item The behavior of the thermal variables and dynamic variables are
affected by different aspects of the models and numerical scheme.
For example, the fluctuations
of density and pressure are dominantly determined by physical models while
the fluctuations of velocity are sensitively dependent on
numerical parameters, e.g., $c_\mu$ and $C_2$.
  \item There is no constant ratio of $w''_z/(w''_x+w''_y)$ for
anisotropic turbulence in current models. We suggest that  $c_3$ take an infinite value at the
boundary and approach its minimum ($0.75$) in the deep convective region.
  \item The diffusive models for non-local transport are not applicable
since the coefficients for different quantities are dramatically
different. The best situation is the turbulent transport of
turbulent kinetic energy where a roughly flat region exists for
$l_1$.
  \item For current resolutions, $c_\mu=0.25$ is better than $c_\mu=0.2$.
And $C_2$ should be set as small as possible in any cases. A
flux-splitting technique is needed to stabilized the shock waves
near the top. A Rayleigh number less than $10^9$ may smear the
turbulent motions.
\end{enumerate}

Our simulations may suffer from the lower resolutions, aspect ratio
and the number of testing cases. But
it is enough for the purpose of validation and get some preliminary results.
A further study will be performed in the future.

\begin{acknowledgements}
We wish to acknowledge the contributions of K. Xu to the developing
of the hydrodynamic code. We also thank the Department of Astronomy
at Peking University for providing computer time on their SGI Altix
330 system. This work was partially funded by the Chinese National
Natural Science Foundation (CNNSF) through 10573022, 10773029 and by
the national 973 program through 2007CB815406. KLC thanks Hong Kong
RGC for support.
\end{acknowledgements}

\label{lastpage}

\end{document}